%% file: main.tex
\documentclass[sigconf]{acmart}
\usepackage{natbib}
\usepackage{balance}
\usepackage{mathrsfs}
\usepackage{amsmath,amsfonts,amsthm}
\usepackage{enumitem}

\usepackage{graphicx}
\usepackage{extarrows}
\usepackage{verbatim}
\usepackage{makecell}
\usepackage{diagbox}
\usepackage{bm}
\usepackage{multirow}
\usepackage{float}
\usepackage{placeins}
\usepackage[linesnumbered, ruled]{algorithm2e}
\SetAlFnt{\small}
\usepackage{epstopdf}
\usepackage{setspace}
\usepackage[super]{nth}
\usepackage{fancyhdr}
\usepackage{slashbox}
\usepackage{subcaption}
\usepackage{lipsum,multicol}
\usepackage{url}
\usepackage{xcolor}
\usepackage{booktabs}
\usepackage{diagbox}
\usepackage{tikz}
\allowdisplaybreaks[4]

\AtBeginDocument{%
  \providecommand\BibTeX{{%
    \normalfont B\kern-0.5em{\scshape i\kern-0.25em b}\kern-0.8em\TeX}}}

\copyrightyear{2023}
\acmYear{2023}
\setcopyright{acmlicensed}\acmConference[MMSys '23]{Proceedings of the 14th
ACM Multimedia Systems Conference}{June 7--10, 2023}{Vancouver, BC,
Canada}
\acmBooktitle{Proceedings of the 14th ACM Multimedia Systems Conference
(MMSys '23), June 7--10, 2023, Vancouver, BC, Canada}
\acmPrice{15.00}
\acmDOI{10.1145/3587819.3590986}
\acmISBN{979-8-4007-0148-1/23/06}

\newcommand{\name}{{\em MOSAIC}} 
\newcommand{\names}{{\em MOSAIC }}
\newcommand{\namef}{{\em MOSAIC's\ }}

\begin{document}

\title{\name: Spatially-Multiplexed Edge AI Optimization over Multiple Concurrent Video Sensing Streams}


\author{Ila Gokarn}
\affiliation{%
    \institution{Singapore Management University}
    \country{Singapore}}
\email{ingokarn.2019@phdcs.smu.edu.sg}

\author{Hemanth Sabella}
\affiliation{%
    \institution{Singapore Management University}
    \country{Singapore}}
\email{hemanthrs@smu.edu.sg}

\author{Yigong Hu}
\affiliation{%
    \institution{University of Illinois at Urbana-Champaign}
    \country{USA}}
\email{yigongh2@illinois.edu}

\author{Tarek Abdelzaher}
\affiliation{%
    \institution{University of Illinois at Urbana-Champaign}
    \country{USA}}
\email{zaher@illinois.edu}

\author{Archan Misra}
\affiliation{%
    \institution{Singapore Management University}
    \country{Singapore}}
\email{archanm@smu.edu.sg}

\renewcommand{\shortauthors}{Gokarn, et al.}

\begin{abstract}
Sustaining high fidelity and high throughput of perception tasks over vision sensor streams on edge devices remains a formidable challenge, especially given the continuing increase in image sizes (e.g., generated by 4K cameras) and complexity of DNN models. One promising approach involves \emph{criticality-aware processing}, where the computation is directed selectively to ``critical" portions of individual image frames. We introduce \name, a novel system for such criticality-aware concurrent processing of multiple vision sensing streams that provides a multiplicative increase in the achievable throughput with negligible loss in perception fidelity. \names determines critical regions from images received from multiple vision sensors and spatially bin-packs these regions using a novel multi-scale Mosaic Across Scales (MoS) tiling strategy into a single `canvas frame’, sized such that the edge device can retain sufficiently high processing throughput.  Experimental studies using benchmark datasets for two tasks,  Automatic License Plate Recognition and Drone-based Pedestrian Detection, shows that \name, executing on a Jetson TX2 edge device, can provide dramatic gains in the  throughput vs. fidelity tradeoff. For instance, for drone-based pedestrian detection, for a batch size of 4, \names can pack input frames from 6 cameras to achieve (a) $4.75\times$ ($475\%$) higher throughput (23 FPS \emph{per camera}, cumulatively 138FPS) with $\leq1$\% accuracy loss, compared to a First Come First Serve (FCFS) processing paradigm.
\end{abstract}



\maketitle
\input{01-introduction}
\input{02-first_principles}
\input{03-overview}

\input{04-system_implementation}
\input{05-evaluation}
\input{06-discussion}
\input{07-related}
\input{08-conclusion}

\begin{acks}
This work was supported by National Research Foundation, Singapore under its NRF Investigatorship grant (NRF-NRFI05-2019-0007), and in part by The Boeing Company, IBM (IIDAI), ARL W911NF-17-2-0196, and NSF CNS 20-38817. Any opinions, findings and conclusions or recommendations expressed in this material are those of the author(s) and do not reflect the views of National Research Foundation, Singapore.
\end{acks}

\bibliographystyle{ACM-Reference-Format}
\bibliography{biblib_ila}

\end{document}

%% file: 01-introduction.tex
\section{Introduction}
\label{sec: intro}
A growing number of real-time applications of machine perception (e.g., 3D scene analysis for autonomous driving or vehicular tracking by street-mounted cameras) involves the execution of DNN-based inference over \emph{multiple, concurrent, high-resolution} multimedia sensor data streams on a resource-constrained edge device. Real-time edge-based execution of such perception tasks remains challenging, given the rapid increase in both DNN model complexity and data size/resolution (e.g., 4K image frames). For example, an NVIDIA Jetson TX2~\cite{xavier} device can process a  maximum of only $\sim$2 frames per second (FPS)  when executing the YOLOv5L6 (191 layers, 47M parameters) object detector at FP16 precision on a $1280\times1280$ image frame. Conventional approaches for overcoming this throughput/latency challenge include either (a) the use of smaller, less accurate DNN models executing on lower resolution data (e.g., a less complex $300\times300$ SSD model\cite{liu2016ssd} can achieve a processing throughput of 10-15 FPS on the TX2 with TensorRT optimizations)  or (b) the use of more expensive, higher-resourced edge devices (e.g., the Jetson Xavier\cite{xavier}). Such approaches impose an unfavorable \emph{cost/complexity vs. accuracy} tradeoff.

\begin{figure*}[thb]
  \begin{minipage}{.37\textwidth}
    \includegraphics[height=2in]{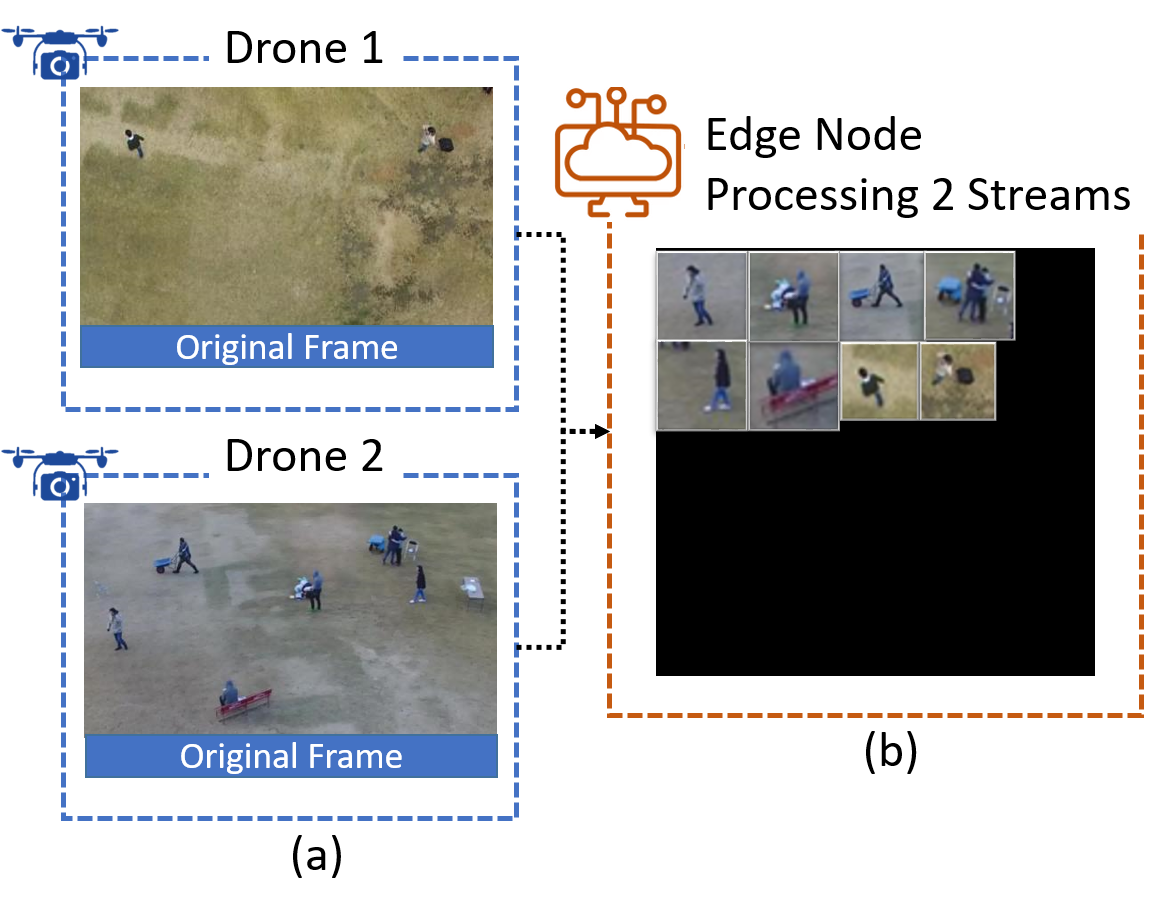}
    \vspace{-0.1in}
    \caption{\namef Overall Functionality: (a) Input frames captured by cameras (b)  Packing tiles from multiple images onto a single canvas frame (image not to scale).}
    \label{fig:canvas-bigpicture}
  \end{minipage} \quad
  \begin{minipage}{.3\textwidth}
    \centering
    \includegraphics[height=2in]{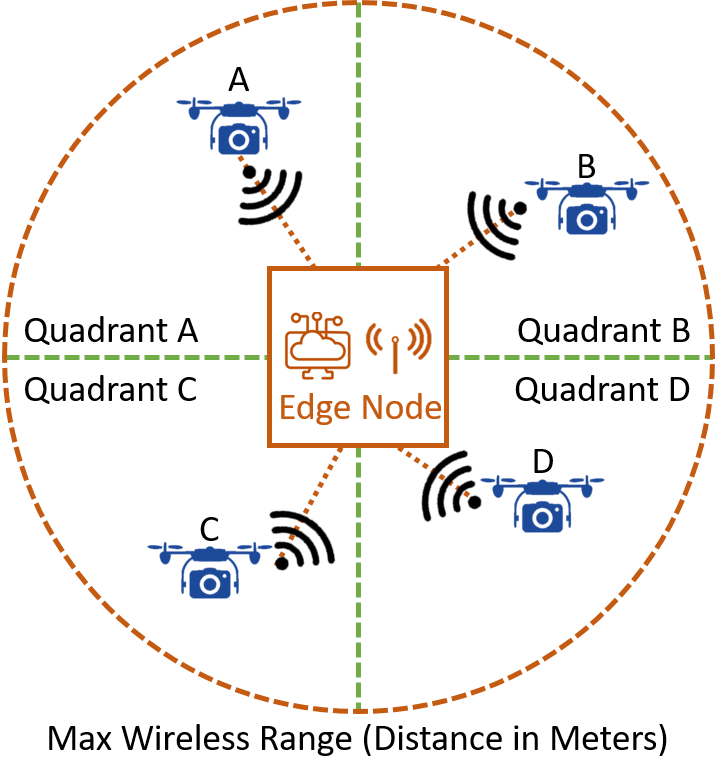}
    \vspace{-0.1in}
    \caption{Motivating Application: Multiple discrete autonomous drone camera streams transferred wirelessly to, and processed by a single edge node.}
    \label{fig:target-app}
  \end{minipage} \quad
  \begin{minipage}{.28\textwidth}
    \centering
    \includegraphics[width=1.9in]{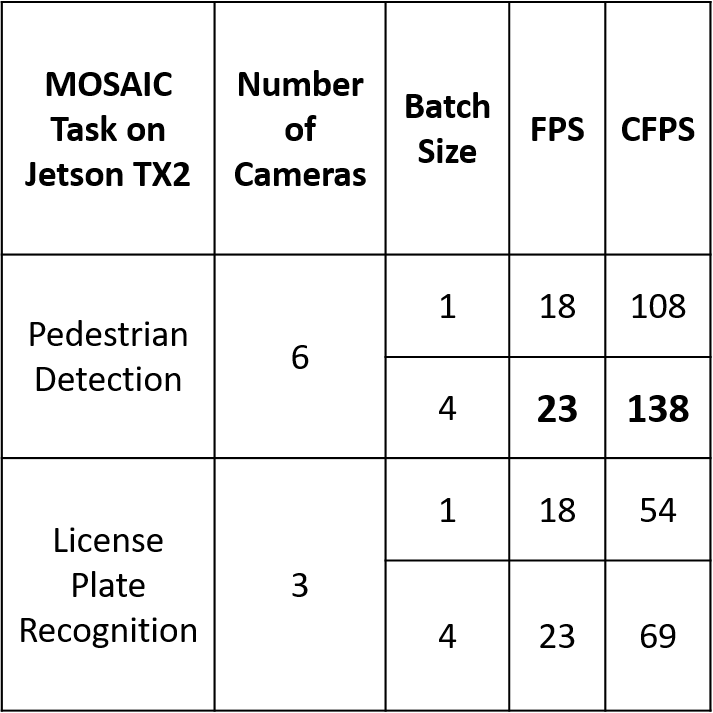}
    \caption{\names System Performance on Jetson TX2; CFPS = Cumulative FPS across M cameras}
    \label{tab:fps-table}
  \end{minipage}
  \vspace{-0.1in}
\end{figure*}

To reduce this computational overhead, recent works have proposed either (a) \emph{criticality-aware processing} approaches, where only selective high-value portions of individual image frames are processed with higher attention or fidelity~\cite{gao:cvpr2018,yi:mobicom2020,yang:infocom2022} or offloaded for DNN task inference~\cite{zhang2021elf} or (b) \emph{selective computation approaches}, where certain DNN layers are simplified~\cite{bhattacharya-sensys16} or skipped~\cite{xin-eccv18}. Current approaches, however, do not consider scenarios where \emph{multiple} sensor streams, with dynamically varying scene characteristics (e.g., object sizes and speeds) share the same computational resources on an edge device and must be processed \emph{concurrently}.


To address this gap, we introduce \name, a criticality-driven edge processing system that optimizes the edge-based execution of DNN-based inferencing tasks over multiple image streams. Through this optimization, \names provides a \textbf{multiplicative increase in the \emph{per-stream throughput}} that an edge device can sustain  \textbf{without sacrificing task accuracy}. \names can benefit a wide variety of applications including, but not limited to, (i) city traffic monitoring, where multiple camera streams monitoring an intersection are processed on a single edge node, and (ii) urban event monitoring, where multiple drones' camera streams are processed by a single handheld edge control unit, as illustrated in Figure~\ref{fig:target-app}.

\namef central concept involves the notion of a \emph{Canvas}, defined intuitively as the maximum size of an input frame (say $C$) that a DNN, executing on a GPU-equipped edge device, can consume while ensuring that the processing throughput remains above a minimal FPS threshold. We call these reduced-resolution frames, which the GPU can keep up with, \emph{canvas frames}. The challenge of concurrently processing multiple (say $M$) camera streams can then be framed as one of spatially \emph{packing} or fitting high-priority regions from $M$ independent image frames into a $C$-sized canvas frame. Conceptually, \names replaces the baseline mode of independent, \emph{sequential} DNN execution on each individual frame with a \emph{spatially-multiplexed} paradigm, where $M$ frames (one from each camera sensor) are processed concurrently. 

\namef design addresses two key challenges with this paradigm: (a) identifying and extracting critical regions from frames with very low overhead, so as to sustain high throughput, and (b) allocating the shared canvas space equitably across critical regions with dynamically varying object characteristics. To pack the canvas frame appropriately, \names efficiently decomposes an input frame into multiple tiles (sub-regions), defined at different scales or ``levels of zoom" that collectively represent objects/RoI of different sizes. The selected tiles that contain faithful representations of the objects or Regions of Interest (RoI) are then \emph{inverse-2D bin packed}~\cite{chung2015notes} onto a canvas frame, thereby providing an $M-fold$ boost in processing throughput as illustrated in Figure~\ref{fig:canvas-bigpicture}(b)). \namef packing is carefully designed to ensure that (a) tiles are proportionally resized based on their criticality while ensuring that the underlying object sizes conform to application-defined spatial bounds, and (b) the tiling process, invoked intermittently, is very low-overhead.

Via benchmark datasets for two distinct applications - \emph{Okutama-Action}~\cite{barekatain2017okutama} for drone-based pedestrian detection and \emph{UFPR-ALPR}~\cite{laroca2018robust} for license plate recognition (\emph{LPR}), we demonstrate \namef ability to significantly improve the throughput-vs.-accuracy tradeoff for diverse machine perception tasks across diverse camera settings.

\noindent \textbf{Key Contributions:} We make the following key contributions:
\begin{itemize}[leftmargin=*]
\item \emph{Criticality-Preserving Canvas-Based Processing:} We develop a 3-stage innovative pipeline, called Mosaic Across Scales (MoS), to dynamically fit a variable number of critical regions, from multiple camera input images, into a single canvas frame: (i) a multi-scale tiling mechanism that uses a Min-Cost Min Set Cover algorithm~\cite{alon2009online} to select an appropriate \emph{minimal} subset of all possible tiles, for any given camera input image, that both capture all likely objects of interest while assuring such objects the largest area possible in the eventual canvas frame, (ii) a Min-Max optimization technique to differentially resize such individual selected tiles based on their computed criticality values, and (iii) a Differential Evolution Algorithm~\cite{mysticlib} with geometric constraints for 2D Inverse Bin Packing~\cite{chung2015notes} all selected tiles (across $M$ distinct images) onto a canvas frame. For pedestrian detection, MoS suffers a negligible ($\leq1\%$) accuracy loss when compared to the low-throughput, sequential FCFS processing of images; it also achieves an $8\%$ increase in accuracy when compared to a strategy of uniformly downsampling and packing all $M=6$ images onto a single canvas frame. For LPR, \names can pack tiles containing vehicles differentially from $M=3$ images to achieve Optical Character Recognition (OCR) Character Error Rate (CER) of <$33\%$, far superior to a baseline uniform resizing approach, that results in $100\%$ CER (complete loss of readability) even when frames are processed individually ($M=1$) i.e. FCFS processing of each input frame at the same DNN inference image size (or canvas frame size) as \names.

\item \emph{Real-time \names Implementation and Performance Gains: } We implement \names using the NVIDIA Jetson TX2 as the edge device. The developed \names system uses a combination of intermittent full-frame object localization and continuous motion tracking  to support low-overhead, approximate extraction of critical regions. Table~\ref{tab:fps-table} details the throughput for packing images from $M=$6 cameras for pedestrian detection, and $M=3$ images for LPR. For both the pedestrian detection and LPR applications, \names achieves 5x (batch size=1, 18 FPS per camera) and 4.75x (batch size=4, 23 FPS per camera) increase in processing throughput, compared to an FCFS baseline that processes each input frame sequentially. 
\end{itemize}

%% file: 02-first_principles.tex
\begin{figure}[thb]
  \begin{minipage}{.48\linewidth}
    \centering
    \includegraphics[width=\linewidth, height=1.6in]{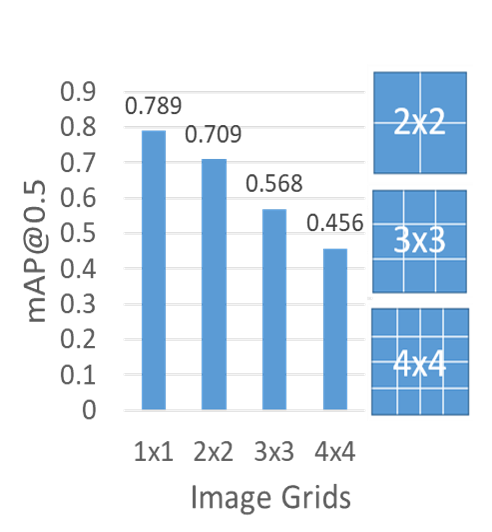}
    \vspace{-0.2in}
    \caption{Accuracy vs. Grid Size ( Uniformly Packing)}
    \label{fig:fp-grid}
  \end{minipage} 
  \hfill
  \begin{minipage}{.48\linewidth}
    \centering
    \includegraphics[width=\linewidth, height=1.6in]{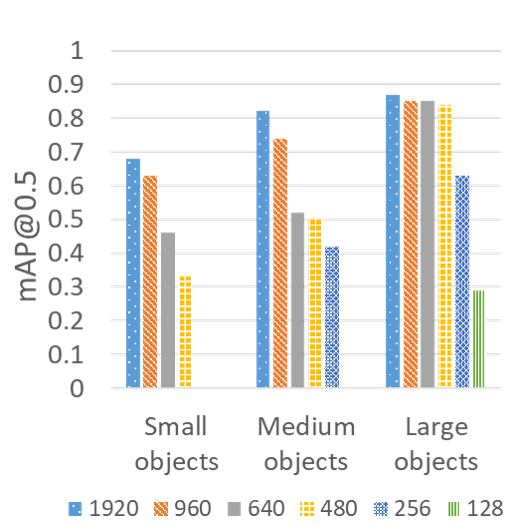}
    \vspace{-0.2in}
    \caption{Object Detection Accuracy vs. \{Object size, Resolution\}}
    \label{fig:fp-size}
  \end{minipage} \quad
  \vspace{-0.2in}
\end{figure}

\section{Motivating \name}
\label{sec: background}
\namef primary objective is to pack high-priority or critical regions from $M$ discrete camera streams onto a single canvas-frame to increase frame processing throughput for each stream, without sacrificing inference task accuracy. To this end, we first explore the target applications and basic principles that underpin \namef key design decisions. 

\begin{figure*}[thb]
        \centering
        \vspace{-0.1in}
        \includegraphics[width=0.8\textwidth]{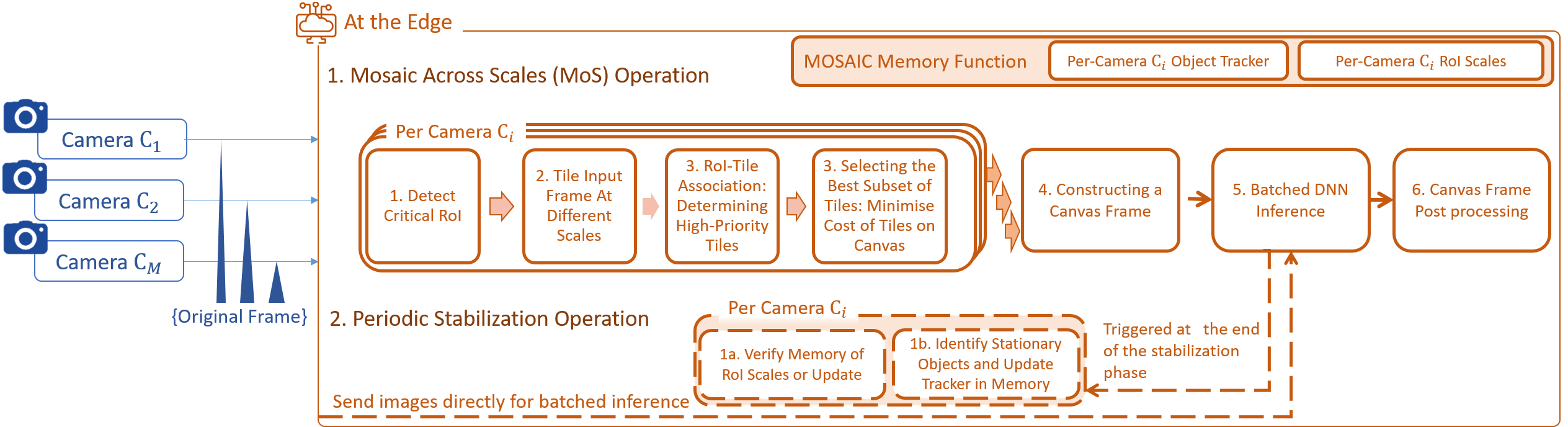}
        \caption{\names Block Diagram of Sub-Components Operating at the Edge}
        \label{fig:blockdiagram}
        \vspace{-0.1in}
\end{figure*}

\subsection{Target Applications}
We envision a wide variety of applications that could benefit from the increased throughput afforded by canvas-based processing--examples include surveillance, counting or detection scenarios where both the camera sensor and the edge node could be stationary or in motion. Consider a city traffic monitoring application where an edge node deployed on a road-side unit processes traffic-light camera feeds from $M$ discrete intersections, each observing unique traffic volumes at different times of the day. Alternatively, consider an edge node deployed on a mobile robot or handheld control unit that processes autonomous drone camera feeds for aerial monitoring in search-and-rescue, wildlife poaching or urban crowd control applications. For all of these scenarios, we can reduce infrastructure costs and shorten response times without compromising perception fidelity by allocating computing resources \emph{non-uniformly} and \emph{selectively}, to only relevant portions of the captured  image frames. We assume that each camera sensor monitors a distinct, non-overlapping physical region, although \names can likely be further optimized to take advantage of any spatial overlaps (Section~\ref{sec:discussion}).

\subsection{First Principles}
\subsubsection{Increasing Throughput by Packing Multiple Images}
We first study the implications of spatially packing multiple input image frames uniformly (without any criticality awareness) into an image grid, as a means of increasing processing throughput. Figure~\ref{fig:fp-grid} plots the object detection accuracy (computed as \emph{mAP} or mean average precision), as a function of the number of such image frames packed, for frames from the Okutama-Action\cite{barekatain2017okutama} that have a native resolution of $3840\times2160$. For all studies in this work, canvas size is set as $640\times640$, adopted from reported benchmarks for YOLOv5~\cite{yolov5}, the model of choice in our work. As seen in Figure~\ref{fig:fp-grid}, the higher the number of elements (distinct frames) in the grid, the smaller the resulting region (pixels) allotted to each input frame, and the lower the object detection accuracy. Intuitively, due to uniform downsizing, smaller objects become progressively smaller and less distinguishable, to the point of loss of  detect-ability. There is therefore an opportunity for the system to increase inference accuracy by providing high-priority regions of the original frame a larger spatial share of the canvas. 

\subsubsection{Factors Impacting Object Detection Confidence}
To further delve into why packing an image into a smaller sized grid (i.e., reducing its overall pixel count) results in lower detection accuracy, we additionally analyze the variation in object detection accuracy over different object sizes. We experimentally observe that object detection accuracy values degrade either due to a reduction in an object's size (a natural consequence of downsizing an image frame to fit into a smaller grid) or a loss in object resolution (greater pixelation). As shown in Figure \ref{fig:fp-size}, as image sizes increase, object detectability increases leading to an increase in average object object detection accuracy. Overall, image downsizing has a \emph{variable} impact: small and medium sized objects stand to receive the largest confidence boost from the increase in resolution and size, while the detectability of large objects remains relatively robust to resolution loss. 

\subsection{\namef Design Choices}
At the edge, \names dispenses with the straightforward approach of using uniform downsizing to pack multiple image frames into a single canvas frame. Instead, \names seeks a differential downsizing strategy, which seeks to \emph{reduce the disparity in pixel areas} corresponding to likely objects (varying in size and spanning multiple input images) embedded in the canvas frame prior to DNN inference. To reduce such disparity (which improves the accuracy for smaller objects without disproportionately penalized larger objects), \names first uses variable-sized tiles to optimally capture pixel regions with likely objects, and then spatially resizes such tiles within acceptable bounds to fit within the target canvas frame. 

%% file: 03-overview.tex
\section{\name}
\label{sec:mos}
We now describe the criticality-aware adaptive processing performed by \names at the edge, as illustrated in Figure~\ref{fig:blockdiagram}. \names alternates between two modes of operation (1) Mosaic Across Scales (\textbf{MoS}) and (2) Periodic Stabilization (\textbf{PS}), both of which interact with \namef Memory Function to support \names objectives: (i) extraction of critical regions from an input frame and (ii) bin-packing of these critical regions into a canvas frame for DNN inference. 
The Periodic Stabilization (PS) operation initialises and intermittently refreshes \names pipeline by running full-frame batched DNN inference on all camera streams to detect class-specific critical objects. For each camera, PS localises the newly detected objects, identifies stationary objects, and updates the object tracker maintained in \namef Memory Function to correct all tracker uncertainties accumulated from the previous round of MoS operation. PS also prompts \names to examine the recently observed RoI and object size distributions per camera to compute a set of camera-specific RoI scales. These RoI scales are updated in the Memory Function and are used to instruct the next round of MoS operation on the sizes of expected RoI in each camera stream. 
For each camera stream, the MoS operation first estimates the locations of critical RoI with motion-based background subtraction and updates the camera-specific object tracker. MoS extends the philosophy of criticality by tiling the input image at each identified RoI scale, identifying the tiles that contain all the RoIs maintained by the tracker, and resizing the tiles (and by extension the RoI contained within the tile) to larger dimensions as much as possible to boost object detection accuracy. This approach effectively ``spatially channels" the limited computation resources available at the edge to the critical regions. Of course, such resizing is a zero-sum game overall that should preferentially enlarge smaller, distant RoI; this objective is complicated by the reality that a single tile in an input frame can contain RoI of different sizes (e.g., a mix of nearer and distant objects). 

At a high-level, the MoS process must balance two conflicting objectives: (a) the RoI should ideally consume a large-enough fraction of a tile such that tile resizing does not eventually result in a dramatically smaller object--\emph{this criterion favors smaller tiles}, and (b) the total number of tiles to be fitted into a canvas should ideally be minimized, so as to allow each tile a larger share of the canvas--\emph{this criterion favors larger tile sizes}.  MoS creates a canvas frame through a number of innovative sequential steps, described next, that collectively balance these two objectives by:
\begin{itemize}[leftmargin=*]
\item First, extracting a minimal set of tiles (at each of the camera specific RoI scales maintained in the Memory Function) within each image frame to encompass the likely ROI maintained by the tracker
\item Then, 2-D bin ``inverse" packing such multi-scale tiles (i.e., ensuring all relevant tiles are packed) to construct a canvas frame as a composite image of such tiles; this canvas frame is then sent to the DNN model for inference. 
\end{itemize}

\subsection{Periodic Stabilization (PS)}
The PS operation initializes and refreshes the entire \names pipeline with class-specific objects, their locations, and object size distributions evaluated through batched full-frame DNN inference on all the incoming camera streams for the entire PS duration. The objective of the PS operation is to address two significant challenges faced during MoS operation. 
First, cameras may observe a variety of object size distributions based on the physical installation of the camera and the observer-object distance. For example, a camera mounted on a lamp-post may observe a mix of large foreground and smaller background objects of interest, whereas a camera mounted on a drone may observe uniformly small object distributions. Object size distributions may also be dynamic over time, for example, a camera mounted on a drone may observe varying object sizes as it increases/decreases its flying altitude. To best capture critical RoI of different sizes, MoS tiles the input frame at different scales and evaluates which subset of tiles contain critical RoI from the input frame. MoS relies on the object size distribution evaluated during the PS operation to understand how many scales to use for such a tiling step and what tiling dimensions each scale must adopt. 
Second, the MoS pipeline estimates critical RoI by performing background subtraction, which captures the RoI where objects are likely in motion and may miss stationary, halted, or occluded objects of interest. MoS similarly relies on the detected object locations obtained by the PS operation to update a camera-specific object tracker with the locations of objects that might be missed by the background subtraction based estimation of critical RoI in the input frame.  
The periodicity and duration of PS is a configurable parameter and describes the expected average rate of change in the observed object size distribution. However, the PS periodicity parameter and overall \names achievable throughput is inversely related: a shorter PS and longer MoS period provides larger throughput gain, as the PS period effectively processes frames at full resolution (without any spatial multiplexing gain).  
After completing the full frame detections for the PS duration across all camera streams, \names is triggered to refresh the per-camera ROI scales and object tracker, described next.

\noindent \textbf{(1a) Calculate Per-Camera RoI Scales:}  
To determine the camera-specific RoI scales and their dimensions, this function first collects the object size distribution observed during the PS operation and assesses if any of the detected objects are overlapping with each other or are in close proximity. Such objects may be detected as a single RoI during the MoS operation so this function adds the minimum enclosing rectangle for such overlapping/nearby objects to the observed object distribution. The resulting object size distribution is then clustered using a KNN clustering model, with an elbow-detection method to determine the appropriate value of $k$ (the number of distinct clusters). This $k$ value then determines the number of scales that MoS will employ for that specific camera, with the centroid of each of these $k$ clusters determining the size of the corresponding tile. MoS takes the larger value between each centroid's x and y coordinate for each scale i.e. $max(x_{centroid}, y_{centroid})$ as the tiling dimension for that scale, rounded to the nearest multiple of 32 for computational efficiency, to best represent objects whose size falls within this cluster. 

Figure~\ref{fig:mos-tiling} illustrates this process by plotting the object distribution observed over all frames by Camera 1 from the Okutama-Action dataset; clustering identifies 3 clusters with centroids $(36, 39)$, $(50, 54)$ and $(81, 44)$ respectively. MoS consequently determines the 2 scales of tiles to be $64\times64$, and $96\times96$, respectively. MoS also introduces a \emph{catch-all} tile, $\sim1.5x$ larger than the largest determined tile (in our example, this results in an additional $128\times128$ tile), to accommodate the possibility of subsequently observing objects/RoI larger than anything seen during the preceding PS operation. 

\noindent \textbf{(1b) Identify Stationary Objects and Update Per-Camera Tracker in Memory Function:} 
This function initializes and refreshes a per-camera Kalman Filter Centroid-based tracker which maintains the most recently observed location and state for all objects in the camera FoV. This state refers to whether the object is ``active" (i.e. in motion) or ``stationary" and is computed by observing the distance travelled by each centroid during the entire PS duration. 

\subsection{Mosaic Across Scales (MoS)}
MoS begins operation on the camera frames ingested immediately after the PS operation completes. As seen in Figure~\ref{fig:blockdiagram}, Steps 1 through to 4 in the MoS operation are carried out on each camera stream in parallel until the relevant tiles containing critical RoI from input frames across \emph{all} camera streams are determined. These relevant tiles are then bin-packed into a canvas frame for DNN inference and post-processing. 

\noindent \textbf{(1) Determining Critical Regions of the Input Image:} 
This MoS sub-component assembles a list of mask bounding boxes where critical RoI are estimated to be present in the input frame. To achieve this, the ingested input frame is compared to the previous frame for background subtraction which detects critical RoI that might contain objects in motion. MoS updates the mask bounding boxes and the camera object tracker with these detected RoI locations for matching with known RoI tracks and assigns the status of the updated tracks as ``active". This method is robust to new objects that enter or existing objects that move in the camera Field of View (FoV). Among the tracks not updated by the RoI from the current input frame, MoS assesses the status of each track. For tracks marked ``stationary" and missed by the background subtraction-based motion estimation, MoS retrieves their last known locations from the tracker memory to add to the mask bounding boxes. For the remaining unmatched tracks, MoS assumes that the object that was previously in motion has either come to rest, crossed paths and jumped tracks with another object, or occluded by another object and therefore merged with the other RoI. In all these cases, MoS assigns the status of the track as ``last-seen" and includes all the unmatched track locations to the list of mask bounding boxes for the current input frame. Such ``last-seen" objects are maintained in memory until they are reactivated or until the end of the MoS operation period. This is done to avoid missing objects that might have come to a halt at its last known location for the remaining duration of MoS operation. Lastly, in the event of detected camera ego-motion (e.g., for a camera mounted on a moving drone), this sub-component also performs camera motion compensation which detects frame keypoints, matches the descriptors of the current and previous input frame, and calculates the new location of all known tracks in the current input frame by modelling the camera motion as a partial 2D affine transformation. 

\noindent \textbf{(2) Tiling an Input Image:} MoS then generates a bag of tiles, at each scale dimension maintained by \names Memory Function with a configurable overlap parameter that determines the tile strides.   

\begin{figure}
    \vspace{-0.1in}
  \begin{minipage}{.45\columnwidth}
    \centering
    \includegraphics[width = 1.45in, height=1.45in]{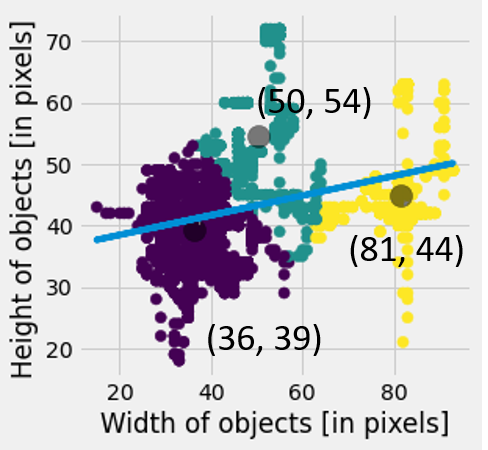}
    \caption{Object Size Distribution \& Clusters in Okutama-Action Drone Sequence 1.1.8}%
    \label{fig:mos-tiling}
  \end{minipage} \quad
  \begin{minipage}{.45\columnwidth}
    \includegraphics[width = 1.45in, height=1.45in]{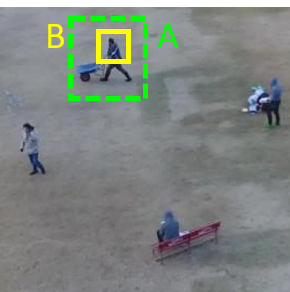}
    \caption{Tiles at different scales, capturing objects with different ``goodness"}%
    \label{fig:mos-subsets}
  \end{minipage}
  \vspace{-0.1in}
\end{figure}


\noindent \textbf{(3) Determining High-Priority Tiles:} The generated bag of tiles at $k$ different scales may have no objects/RoIs, partial views of objects, or completely contained objects. MoS next determines the subset of such tiles that adequately capture the critical regions, while balancing the two conflicting objectives mentioned earlier. MoS constructs a spatial quadtree from all the generated tiles in the bag of tiles and then uses the assembled mask bounding boxes from Step 1 to perform an intersecting bounding box search. All tiles that intersect with each of the masks are then evaluated for ``goodness of fit of the mask'' in the individual tile. Figure~\ref{fig:mos-subsets} illustrates such a selection, where MoS chooses tile (A) over tile (B), as tile B only partially captures the human object. 

More formally, MoS assigns a mask to a tile if it satisfies two distinct ``goodness" criteria. First, the tile must capture a significant portion ($\geq 95\%$) of the mask, both in width and height; this ensures that the object is sufficiently visible (and not unduly cropped) to be successfully detected by the downstream object detector DNN. Second, the mask:tile height ratio must lie in the range $(0.5, 0.9)$--i.e., the object must not encompass either too small or too large a fraction of the tile's area. The lower bound increases the probability that the object will retain larger dimensions after being resized and 2D bin-packed onto the canvas; masks smaller than half the tile's dimensions are effectively assigned to tiles of smaller scales. Conversely, the upper bound (0.9) ensures that relatively large-sized objects are preferentially assigned to tiles of larger scale (which can likely accommodate additional objects), while also minimizing the risk of undue cropping due to an inaccurate mask. Via this process, MoS curates a filtered bag of tiles, each containing faithful representations of estimated objects at the appropriate scale, thereby assuring that such objects will remain reasonably sized (with higher likelihood of successful DNN detection) when the canvas frame is composed. 


\noindent \textbf{(4) Selecting The Best Subset of Tiles:} A number of different combinations of tiles might ``cover" (in a set-theoretic sense) all masks/objects that need to be included onto the canvas for inference. However, to promote efficient (less-redundant) packing onto the canvas frame, it is imperative to select only those tiles that are not only likely to preserve object dimensions in the canvas but also that minimize `wasted pixels' (intuitively, the total number of pixels representing the background or other irrelevant objects, as well as objects captured in multiple tiles). This dual optimization process can be conceptualized as bin-in-a-bin packing problem, where MoS must not only ensure that all objects are `covered' by the chosen tiles, but also that the chosen tiles generate the lowest count of `wasted pixels' possible. We perform such selection by using a greedy approximation (due to the problem being intrinsically NP-Hard) of the Min-Cost Min-Set Cover (\emph{MCMSC}) Algorithm summarized in Algorithm \ref{mainalgo}. 

\begin{algorithm} 
 \small
\SetAlgoLined
\KwResult{Subset of chosen tiles $S$}
$Universe = {m_1, m_2, ... m_m}$ Set of M masks\;
$Tiles = {t_1, t_2, ... t_n}$ Set of N tiles that may contain one or more assigned masks\;
$Costs = {c_1, c_2, ... c_n}$ Set of costs for N tiles\; 
$S \gets \emptyset$\; 
 \While{$S != Universe$}{
  \If{$\mid t_i - S\mid \geq 0 \And (c_i / \mid t_i - S\mid) > 0$}{
  $Subset \gets \min(c_i/ \mid t_i - S\mid) $\; minimize the number of tiles containing the same mask and minimize the additional cost to the canvas associated with adding an additional tile \newline
  $S \gets S \cup Subset$\;
  }}
 \caption{Greedy Min Cost Min Set Cover Algorithm}
 \label{mainalgo}
\end{algorithm}

Intuitively, the MCMSC algorithm selects those tiles that together minimize the cost of wasted background/non-object pixels appearing in the tile while ensuring that each object of interest is part of at least one tile that satisfies the goodness criteria mentioned earlier. To achieve this goal, MoS assembles two distinct views of the mask-to-tile relationships in the filtered bag of tiles. In the first view, for each mask, MoS assembles a set of acceptable tiles that best capture that mask. To calculate the the cost of including the tile into the canvas, the second view concurrently consolidates all the masks assigned to an individual tile. A Min Set Cover over the first view ensures that all masks (possible object-related pixels) are included in the canvas, while a Min Cost over the second view selects the minimal subset of tiles that the canvas must accommodate. 

This unique formulation has several advantages. First, objects that can feasibly be mapped to multiple tiles typically appear in only one (or at most two) tiles in the eventual canvas, reducing the likelihood of false positives in the post-DNN non-maximal suppression (NMS) based inference step.  Second, multiple objects that occur in close spatial proximity are usually represented by a single larger-scale tile (with reduced wasted pixels), instead of being represented by multiple individual smaller-scale tiles. In particular, the min set cover step (line $7$) in the optimization chooses the tile containing the most number of masks not already present in the canvas, while the min cost step (also in line $7$) takes into account the cardinality or the unique number of masks added to the canvas by a single tile and the associated cost of including the tile in the canvas. At the end of this step, MoS computes the final, optimal subset of tiles for each individual camera sensor frame. For each tile in the chosen subset, MoS also computes a spatial sizing bound and an \emph{elasticity factor} (based on the combination of object/tile's scale and application requirements), as the range and amount of resize that can be tolerated during canvas frame construction. We empirically observe that this size bound is \emph{scale-dependent}: tiles of different scales can tolerate different ranges of resizing,  outside of which objects either become intolerably small or suffer from excessive pixellation on enlargement, severely impacting DNN inference accuracy. Small objects are, in fact, especially sensitive to drastic differences in spatial resizing. 


\noindent \textbf{(5) Constructing a Canvas Frame:} 
After the previous step, MoS has effectively curated a set of tiles, their spatial sizing bounds, and elasticity factor, say $\mathcal{ST}_i$, with heterogeneous dimensions for each input image frame ($F_i$). To now pack sensor data from multiple image sensors onto a canvas frame of a given dimension, MoS needs to determine the modified dimensions of all tiles across all of these $M$ subsets--i.e., $\forall \mbox{\; tile\; } t: \mbox{such that } t \in \bigcup_{k=1}^M \mathcal{ST}_k$ such that they can be packed onto a single canvas frame with defined dimensions. This can be generalised as an Inverse Bin Packing Problem~\cite{chung2015notes} where given a defined set of items and bins, the algorithm must converge on the minimum perturbation to the item-size vector such that all the items can be packed into the prescribed number of bins.

MoS approximates such an optimization by using a \emph{computationally-fast} Differential Evolution Algorithm~\cite{mysticlib} (a form of meta-heuristic optimization) with a Min-Max Optimization objective function that minimises the largest dimensions obtainable within its defined bounds such that the combined area of all included tiles is less than the canvas frame area. The optimizer also takes in the elasticity factor as the tile weight, and spatial sizing bounds limiting the size of each individual tile. The Differential Evolution Algorithm also takes an Equality Penalty function which monitors if the number of packed tiles is lower than the number selected for packing by MoS. If so, the next generation of tiles is required to monotonically decrease or ``squeeze" the size of each tile (based on the elasticity factor) to pack all the tiles onto the canvas frame, while ensuring that the amount of ``squeeze" does not violate the tile's defined lower spatial bounds.  In essence, the solver maximises each tile's size within its acceptable bound, while attempting to 2D bin pack \emph{all} the tiles onto the canvas frame. In the event that all tiles are ``squeezed" to their individual minimum permissible size, MoS relaxes the lower bound and notifies the user of a possible loss of accuracy with the advice to assign fewer camera streams to the edge node (a form of ``admission control"). Upon solver convergence, MoS obtains not only the resized \emph{dimensions} for each tile, but also the \emph{canvas position} where it must be packed for optimal fit.

\noindent \textbf{(6) Postprocessing:} As mentioned earlier, the DNN then executes the inference task on the resulting canvas frame, consisting of the re-packed, re-positioned tiles. Finally, MoS also maintains the tile$\rightarrow$bin spatial mapping for each input image included in the canvas, and uses this to perform post-inference coordinate translation of all detected objects (and their locations) back to the original input frame. After translation, MoS also executes a general Non-Maximal Suppression (NMS) step on the translated boxes for each original input frame to remove any double-counting for objects that might have appeared more than once (inside tiles of different scales) on the canvas. Any other downstream vision processing pipeline is applied thereafter on the post-processed objects.

%% file: 04-system_implementation.tex
\section{System \& Evaluation Design}
\label{sec: system}
\begin{figure}[thb]
        \centering
        \vspace{-0.1in}
        \includegraphics[width=\columnwidth]{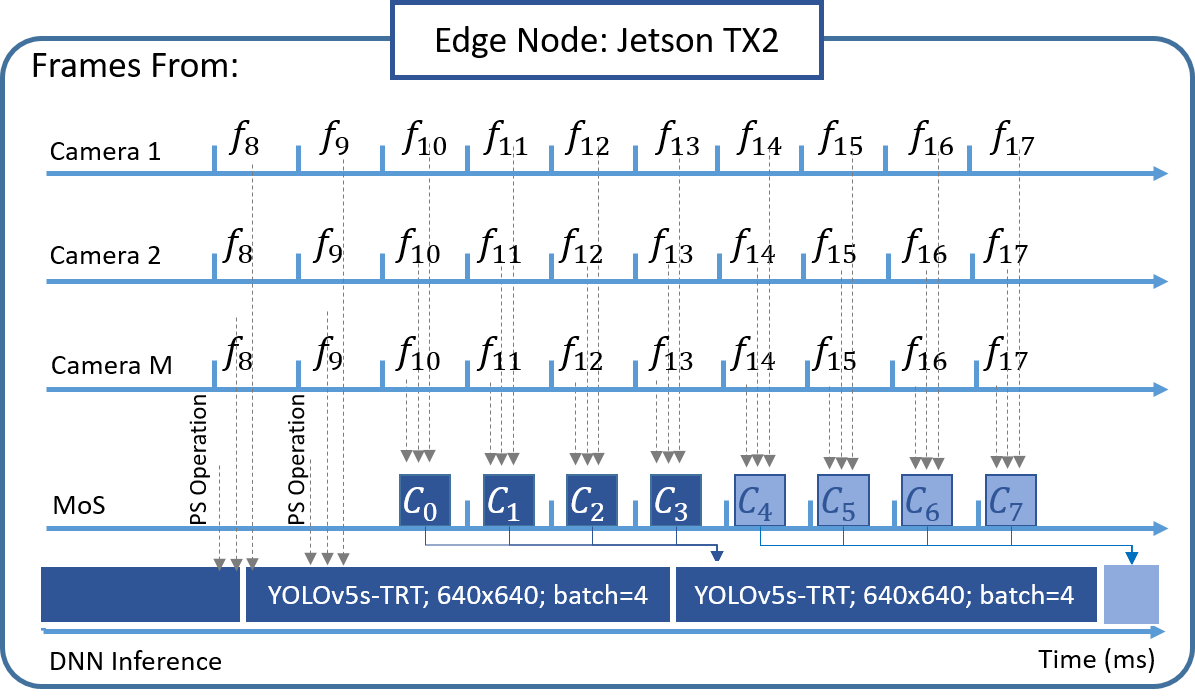}
        \caption{Conceptual design of \namef MoS pipeline at the Jetson TX2 Edge Node}
        \label{fig:rtsys}
  \vspace{-0.1in}
\end{figure}

We deploy the PS and MoS pipelines at the edge, as visualised in Figure~\ref{fig:blockdiagram}. During the PS operation, input frames from $M$ cameras are batched and directly sent for DNN inference. By default, \names executes the PS operation over 10 frames, with a a periodicity of 30 seconds. This setting (see Section~\ref{sec: study2} for a deeper analysis) adequately balances the need for sufficient frames to detect and classify stationary vs. moving objects with the desire to minimally impact \namef overall achievable throughput. During the MoS operation, input frames from $M$ cameras are received and concurrently evaluated by the MoS pipeline to construct a canvas frame, as visualised in Figure~\ref{fig:rtsys}. MoS receives the $i^{th}$ input frames, $f_m^i \forall m \in M$ cameras, and constructs a canvas frame $C_i$ from the chosen subset of tiles across all $M$ frames. The DNN inference task is configured for a batch size of $b$ which allows the pipelined canvas construction of the next $b$ frames $(f_i,\ldots, f_{i+b})$ onto canvas frames $(C_I,\ldots, C_{i+b})$ during DNN inference on the previous batch. The achievable throughput on \names is thus a function of both the PS and MoS modes of operation. We empirically establish that on a Jetson TX2, a TensorRT-optimised YOLOv5s model achieves reasonable accuracy and inference latency of $\sim$ 170msec on $640\times640$-sized canvas frames with batch size=4 (thereby achieving $\approx$ 6FPS per camera or cumulatively 24 FPS).  \names adopts this configuration and determines the maximum value of $M$ or the maximum number of cameras that can be supported at a single edge device for a chosen application and set of camera streams for the MoS operation. This value of $M$ is constrained by two key factors: (i) the canvas construction time for $b$ canvas frames from $M$ camera input frames must not exceed the DNN inference time ($\sim\leq 170ms$) to allow seamless pipelined execution, and (ii) all chosen tiles from $M$ cameras must not violate their application-dependent and scale-dependent spatial sizing bounds when packed onto the canvas frame. \names adheres to both constraints to choose the most appropriate value to $M$ for the application to ensure that \names \emph{always} achieves 24 FPS for \emph{all} $M$ camera input streams during the MoS phase. However, \namef overall achievable throughput is reduced due to the periodic PS operation that processes all incoming camera frames sequentially and without modification. With a batch size $b=4$, the PS operation adds a delay of $10 \times M/24$ seconds for processing 10 stabilization frames across $M$ camera input frames under these default settings. For example, for $M=6$ camera streams, batch size $b=4$, the PS operation adds an overall processing latency of $\sim$2.5 seconds, resulting in an overall achievable \names throughput of 23 FPS across both PS and MoS phases.

\begin{figure*}[thb]
  \begin{minipage}{.50\textwidth}
    \centering
    \includegraphics[height=2.1in]{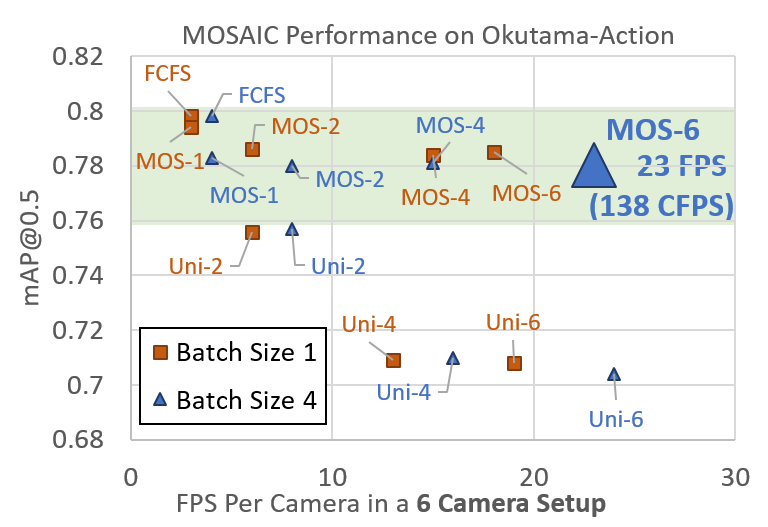}
    \caption{\namef Performance (Pedestrian Detection on Okutama-Action)}
    \label{fig:final-okutama}
  \end{minipage} \hfill
  \begin{minipage}{.49\textwidth}
    \centering
    \includegraphics[height=2.1in]{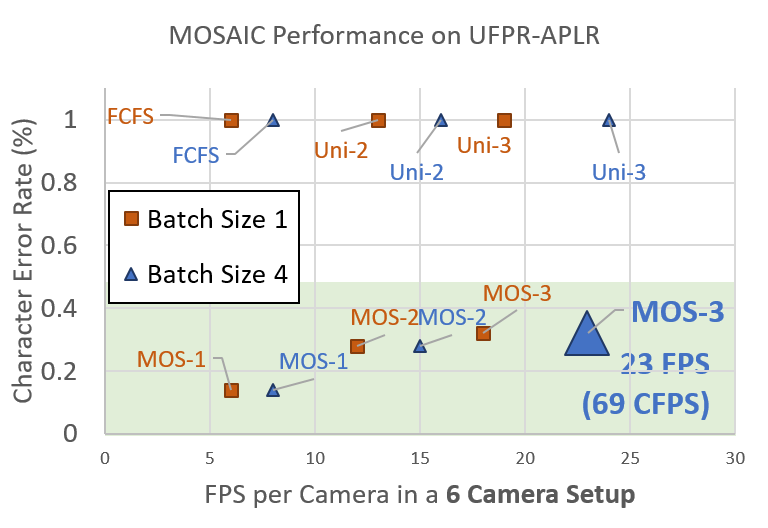}
    \caption{\namef Performance (LPR on UFPR-ALPR)}
    \label{fig:final-ufpr}
  \end{minipage}
  \vspace{-0.1in}
\end{figure*}

\noindent \textbf{Evaluation Platform:} We evaluate {\names} on the NVIDIA Jetson TX2~\cite{tx2}, a representative edge device equipped with a 256 CUDA-core PASCAL GPU, and an ARMv8 multi-processor architecture supporting both a dual-core NVIDIA Denver 2 CPU and a quad-core ARM Cortex A57 MPCore CPU. 


\noindent \textbf{Benchmark Datasets:} We evaluate \names using two benchmark datasets for two distinct applications - \emph{Okutama-Action}~\cite{barekatain2017okutama} for drone-based pedestrian detection and \emph{UFPR-ALPR}~\cite{laroca2018robust} for license plate recognition. The Okutama-Action dataset comprises 43 drone sequences at 4K ($3840\times2160$) resolution encoded at 30FPS, yielding 54664 and 14210 images for training and testing respectively. The UFPR-ALPR dataset similarly comprises 90 video sequences at $1920\times1080$ resolution encoded at 30FPS, yielding 3600 and 1800 frames for training and testing respectively. We consider each video sequence as a distinct camera and use combinations of $M$ cameras without duplication of camera input streams for fair comparisons. \names constructs canvas frames with ``person" objects for the Okutama-Action dataset and vehicle objects of classes \{car, motorcycle, bus\} for the UFPR-ALPR dataset; for UFPR-ALPR, the downstream OCR pipeline then performs LPR on the detected license plate bounding boxes.

\noindent \textbf{Evaluation Model:} For object detection on the canvas frames, we employ a TensorRT-optimised YOLOv5s model, an edge-compatible pretrained model with 7.2M parameters and 16.5 GFLOPs. It is  pre-trained using the MS COCO dataset~\cite{lin2014microsoft} and fine-tuned on the selected datasets for greater sensitivity to occluded, unseen, and small-sized low-resolution objects. 

\noindent \textbf{Evaluation Metrics:} To evaluate possible gains in perception accuracy in the pedestrian detection application, we report the mean average precision of the model at an IoU threshold of 0.5 --i.e. mAP@0.5, and report the inference latency i.e. $FPS_C$ and Cumulative-FPS CFPS = $M\times FPS_C\times b$, where $FPS_C$ is the throughput achieved for canvas frames of size $C$, with batch size $b$, and $M$ cameras per canvas frame. For the license plate detection application, we evaluate the downstream Optical Character Recognition (OCR) quality through Character Error Rate (CER) metric. CER employs the Levenshtein distance metric to calculate the minimum number of single-character changes (i.e. insertions, deletions, or substitutions) required to change the predicted string of characters to the groundtruth, averaged by the number of characters in the groundtruth. The lower the CER rate, the better the OCR performance, with $0$ indicating perfect recognition. 

\noindent \textbf{Evaluation Baselines:} We compare \namef performance against two baselines: 

\noindent \textbf{1. FCFS:} Received frames resized to the canvas frame dimensions and sent for batch DNN inference without any spatial modification--in effect, this is identical to \namef behavior during the PS phase.

\noindent \textbf{2. Uniform-$M$:} Denoted as Uni-$M$ where $M$ signifies the number of images packed onto a single canvas frame. Uniform-$M$ divides a canvas into equal number of grid rows and columns and assigns each input image to a single cell in the grid. Uniform-$M$ also determines which methodology among grid, horizontal, or vertical stacking of $M$ input images creates the best grid structure such that each cell affords its corresponding input frame the lowest possible downsize ratio when compared to its original dimensions. 

%% file: 05-evaluation.tex
\section{Evaluation}
\label{sec: evaluation}
We first evaluate the validity of our fundamental hypotheses, that criticality-aware processing of multiple input frames with our MoS methodology helps improve the throughput vs. accuracy tradeoff for diverse object distributions, camera settings, and applications. We also compare canvas-based processing employed by MoS with batched inferencing methods employed by recent works. For deeper insights into the benefits of MoS, we also conduct ablation studies on how \names performs varies with different parameter settings. 

\begin{figure*}[thb]
  \begin{minipage}{.3\textwidth}
    \centering
    \includegraphics[width=2.2in]{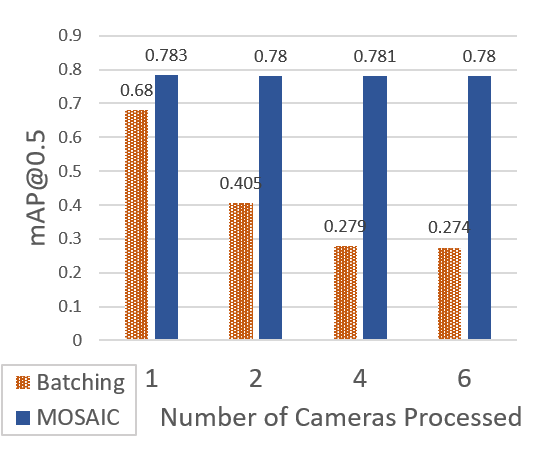}
    \vspace{-0.1in}
    \caption{mAP vs Number of Cameras Processed for Tile Processing Techniques at the Edge}
    \label{fig:batching}
  \end{minipage} \quad
  \begin{minipage}{.3\textwidth}
    \centering
    \includegraphics[height=2in]{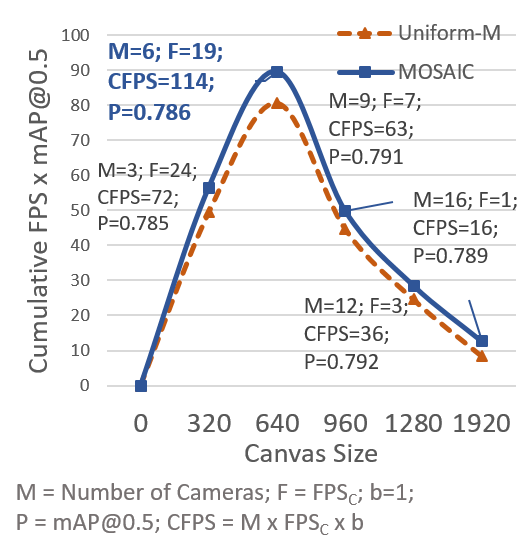}
    \vspace{-0.1in}
    \caption{Canvas Size vs Cumulative FPS x mAP@0.5}
    \label{fig:ablation-mos}
  \end{minipage} \quad
  \begin{minipage}{.3\textwidth}
    \centering
    \includegraphics[width=2.2in]{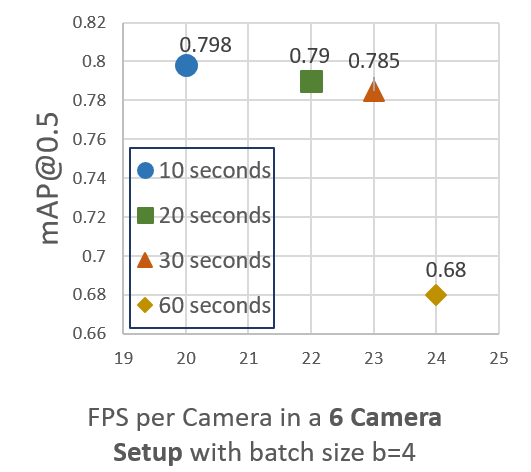}
    \vspace{-0.1in}
    \caption{\namef Performance Gains vs. PS Periodicity}
    \label{fig:ablation-ps}
  \end{minipage}
  \vspace{-0.1in}
\end{figure*}

\subsection{{\names} System Performance}
\label{sec: study5}
In general, we expect throughput and accuracy to be inversely related: increasing $M$ (the number of camera frames being packed into a single canvas frame) should provide an $M$-fold increase in throughput but result in lower mean object detection accuracy. Figures \ref{fig:final-okutama} \& \ref{fig:final-ufpr} plot this accuracy vs. throughput, for different \names configurations and baseline approaches, for $M$ varying between \{1,$\ldots$,6\} for Okutama-Action and \{1,$\ldots$,3\} for UPFR-ALPR, respectively on a canvas frame size $C$ of $640\times640$; the throughput is plotted per camera by dividing the overall processing throughput (i.e. CFPS) by $M$. As detailed in Section ~\ref{sec: system}, the maximum number of cameras $M$ that can be packed onto a canvas frame depends on the canvas construction time, object sizes/scales, and spatial sizing bounds of the resulting subset of tiles. As the objective in UFPR is to eventually perform OCR on the detected license plate objects, it imposes stricter spatial sizing bounds for each tile which limits the number of cameras for this dataset to $M=3$. This is unlike the pedestrian detection application of Okutama-Action which permits more relaxed spatial sizing bounds (even though it contains smaller ($\sim 64\times64$) person class objects), allowing $M=6$ cameras to be successfully packed onto a canvas frame. For batch size $b=1$, \names takes on average $\sim9ms$ and $\sim13ms$ ($\sim36ms$ and $\sim52ms$ for batch size $b=4$) to build a canvas frame from $M=3$ and $M=6$ cameras respectively, well within the inference deadline of 170ms. Both constraints on $M$ thus satisfied, we compare Uniform-$M$ for each application accordingly.  

\noindent \textbf{Pedestrian Detection Application:} In Figure~\ref{fig:final-okutama}, for batch size $b=1$, FCFS (where each image is processed independently and sequentially--i.e., $M=1$) offers the highest accuracy $\sim$0.79 but suffers from very low throughput $\le$3FPS per camera in a 6-camera setting. On the other end of the spectrum, Uniform-$6$ (where $M=6$ images are uniformly compacted into the canvas frame) offers the highest throughput ($\sim$ 19 FPS) per camera, but suffers a significant $8\%$ loss in accuracy to $\sim$ 0.71.  In contrast,  \names offers a significantly more favorable tradeoff as $M$ is varied. With $M=6$: (a) compared to FCFS, MOSAIC-6 experiences only a negligible $\sim < 1\%$ accuracy loss but achieves over 500\% (5x) higher throughput; (b) compared to Uniform-6, MOSAIC-6 achieves significantly higher accuracy $+7.8\%$ with minor 0.04\% reduction in throughput (Uniform-6 = 19 FPS; MOSAIC-6 = 18 FPS). This slight reduction is due to the PS operation running for 10 frames every 30 seconds. Similar throughput-vs.-accuracy tradeoffs can be observed with batch size $b=4$ with the exception that the Uniform-6 and MOSAIC-6 methods both achieve 24 FPS and 23 FPS per camera respectively across 6 cameras (144 and 138 Cumulative FPS), with \namef method achieving significantly higher accuracy by $+8\%$ over Uniform-6. 

\noindent \textbf{License Plate Recognition Application:} In Figure ~\ref{fig:final-ufpr}, for batch size $b=1$, Uniform-1 suffers CER=100\% (complete OCR failure) which indicates that uniformly downsizing a single image to fit onto a $640\times640$ canvas causes severe, catastrophic pixelation of the license plate. On the other hand, MOS-1 retains the pixel resolution of the high-priority vehicle object to completely recover OCR capabilities downstream within a reasonable OCR CER value of 15\%. \names further pushes the envelope on OCR quality downstream by also packing tiles likely containing vehicle RoIs from $M=2$ and $M=3$ images onto a canvas frame to similarly recover OCR ability downstream within reasonable OCR CER bounds (29\% and 33\% respectively), while also achieving a much higher system throughput of 23 FPS per camera for the 3-camera system (Cumulative FPS=69).

\noindent \textbf{Comparative Study with Batched Processing of Individual Objects:} We compare \namef performance against recent priority-aware processing works~\cite{liu2020removing, hu2021exploring, liu2022self} which extract high-priority regions of interest from a camera frame and perform batched inference on the selected regions. Compared with sequential frame inference, batched inference processes multiple inputs at one time to improve parallelism and better utilize the computing capacity of the GPU. However, unlike \namef canvas-based execution where tiles of different sizes can be packed into one canvas frame, batched inference requires all input tile images to be the same size. We therefore compare the mean accuracy from the pedestrian detection application to illustrate the benefits of \namef differential tile resizing strategy in the MoS pipeline at the edge. We first obtain the average execution times of a $640\times640$ canvas frame and batched tile images at different batch size/image size combinations through system profiling of the Jetson TX2. The best tile image size for batching-based execution is then determined online as the largest tile image size to run all input tiles and finish no slower than executing one canvas frame. The curated subset of tiles from MoS are resized with padding to this input image size for batched inference. This way, batched inference achieves the same throughput as \namef canvas-based inference for batch size $b=1$. 

Figure~\ref{fig:batching} visualizes the mAP score of processing $B$=\{1, 2, 4, 6\} input frames at the same time, using the Okutama-Action dataset. We see that both canvas-based execution and batching-based execution achieve good accuracy when the number of input frames is 1. As $M$ increases, the accuracy achieved by batched inference drops, with \namef canvas-based method significantly outperforming (a 3-fold higher accuracy when processing 4 or 6 frames concurrently) the batching-based method as $B$ increases. The accuracy for batching-based execution drops much faster due to the requirement that \emph{all} images must be the same size, which leads to a dramatic loss of accuracy especially for the small objects. On the other hand, the non-uniform sizing supported by the canvas-based method is able better to preserve the accuracy of smaller objects. 

\subsection{Ablation Studies}
\subsubsection{\namef Performance Gains on Different Canvas Sizes}
\label{sec: study1}
Given that the maximum number of cameras, $M$, that can be packed onto a canvas frame is limited by the object distribution observed by the cameras and their spatial sizing bounds, it stands to reason that increasing the canvas frame size $C$ could allow for a higher value of $M$ (thereby increasing throughput) and arguably, higher accuracy if all tiles assume the maximum dimension within their sizing bound. However, larger canvas frame sizes $C\geq 640\times640$ impose greater computation loads and incur higher inference latencies per canvas frame. This increase in canvas frame inference latency reflects in a lower reduced cumulative inference throughput, i.e. CFPS = $M\times FPS_C\times b$, where $FPS_C$ is the throughput achieved for canvas frames of size $C$ with batch size $b$.  Figure~\ref{fig:ablation-mos} illustrates such a  throughput-accuracy tradeoff by plotting a joint "\emph{throughput-accuracy}" metric (defined as Cumulative FPS $\times$ mAP@0.5) vs. canvas frame size for the Okutama-Action dataset. We conclude that \names consistently outperforms the uniform resizing and packing or Uniform-$M$ baseline, regardless of the canvas frame size $C$. We also note that the preferred canvas frame size of $640\times640$ provides the highest  throughput-accuracy gains (15.4\%), compared to all other canvas frame sizes. 

\subsubsection{\namef Performance Gains vs. PS Periodicity}
\label{sec: study2}
We have established that the PS operation periodicity impacts the achievable \names throughput due to the processing latency incurred by the FCFS-based DNN inference on all camera input frames from $M$ cameras. A shorter PS periodicity would yield higher number of FCFS processed frames and therefore more accurate class-specific object detections, while negatively impacting \namef overall achievable throughput. A longer PS periodicity might, conversely, incur higher tracker failure for ``last-seen" and lost or unmatched tracks, in turn creating additional tiles (which may or may not contain objects) that MoS will need to spatially pack, thereby reducing task accuracy. Figure ~\ref{fig:ablation-ps} plots the resulting relationship between PS periodicity, \names throughput, and mAP@0.5 for the Okutama-Action dataset. We see that a PS period=30 secs appropriately balances the dual requirements of high throughput (23 FPS) and high object detection accuracy (78.5\%), while a PS period=60 secs increases throughput by only  1 FPS while suffering a steep ($\sim$ 10\%) mAP drop.  

%% file: 06-discussion.tex
\section{Discussion}
\label{sec:discussion}
\noindent \textbf{Operating under Network Bandwidth or Energy Constraints:} In a real-world system with bandwidth or energy constraints, a compute-capable camera platform can additionally minimize redundancy by also downgrading an image's resolution before transmitting, especially if the edge will use \names to downsize frames during processing. Prior work has shown how criticality-aware techniques for differential downsampling of input frames (e.g., MRIM~\cite{wu2022mrim}), can outperform uniform downsampling by helping preserve object details for edge-based vision inferencing tasks. Such cameras can also implement lightweight multi-object detectors and trackers, such as FastMoT~\cite{yukai_yang_2020_4294717}, to  identify class-specific objects of interest even in challenging environmental conditions such as low-lighting and fog. \namef MoS operation can not only co-exist with such criticality-aware transmission, but in fact can benefit from such on-camera processing by effectively eliminating the intermittent \emph{PS} phase.


\noindent \textbf{Canvas-based Processing of Spatiotemporally Correlated Camera Streams:} 
In the case of spatiotemporally correlated camera streams, multiple objects may be visible from different camera streams at different observer-object distances, angles, and perspectives. With respect to our MoS pipeline, there arises the opportunity to further fine-tune the spatial sizing bounds based on (i) appearance of object across multiple intra-camera and inter-camera frames (ii) motion of object and the need for shorter or longer detection periodicity (i.e. how often an object is included on the canvas frame for inspection). Such modifications to the criticality and application dependant spatial sizing bound can certainly expand the functionality of \namef system to a more diverse range of applications, which we leave for future work. 

%% file: 07-related.tex
\section{Related Work}
\label{sec:related}
Recent works on real-time vision pipelines in cyber physical systems are motivated by both (1) the demonstrated need for edge-based concurrent computation of diverse and high-resolution sensor streams~\cite{abdelzaher2020five}, and (2) the reality that state-of-the-art DNNs for vision tasks are often too complex for efficient execution on  the resource constrained edge~\cite{hu2021exploring}. 


With computation power as the key bottleneck, real-time edge AI has attracted increased academic interest. From the system perspective, earlier-stage works focused on analyzing and understanding the intelligent edge platforms with GPUs~\cite{otterness2020amd}, CUDA scheduling~\cite{olmedo2020dissecting}, and CPU/GPU co-scheduling~\cite{bateni2020co}. Alternatively, from the machine learning model perspective, there have been works on optimizing the flexibility in DNN execution~\cite{bateni2018apnet, kim2020anytimenet}, which could further facilitate their deployment in real-time applications. They essentially modify the DNN execution to support various forms of \emph{preemption}, so that corresponding real-time scheduling algorithms could be applied. More recently, attention scheduling~\cite{liu2020removing, liu2022self, kang2022dnn} was proposed and utilized as a novel data-level optimization and scheduling strategy to enable real-time edge AI, where no modification on the DNN model or the underlying operating system is performed. However, they mostly explored the GPU parallel processing capacity by using task batching. To reduce the average inference latency for processing individual video frames on edge devices, prior Region-of-Interest (ROI) approaches attempt to selectively execute the heavyweight object detector DNN only on selected portions of an incoming image frame, identified via techniques such as background subtraction~\cite{yi:mobicom2020, stone2019towards}, the use of a lower-complexity, `pre-processor' DNN~\cite{gao:cvpr2018} and, most recently, \emph{patch extraction}~\cite{yang:infocom2022}. While these techniques are conceptually similar to our curation of high-priority tiles, they largely focus purely on a single sensor feed as opposed to our approach of spatially packing multiple sensor streams within a single image \emph{canvas}.

With network bandwidth as the key bottleneck, some approaches explore attention and criticality in the context of input filtering by limiting the number of frames transmitted , using techniques such as frame sampling~\cite{chen2015glimpse} and micro classifiers and cascading filters camera~\cite{hsieh2018focus}. More recently, temporal correlation across frames has been leveraged to limit the cameras to transfer only certain regions of the frame~\cite{jiang2018chameleon} or adjust resolutions of objects of interest within a frame~\cite{jiang2021flexible, wu2022mrim}. To tackle the challenge of priority inversion in the AI workflow, recent works (e.g.,~\cite{liu2020removing,kang2022dnn}) explore the scheduling and resolution adjustment of each frame based on both its spatial and temporal criticality. Within the broader vision and AI community, a large body of work addresses the computational overhead of DNNs with techniques for early exit and model selection~\cite{bolukbasi2017adaptive}, model compression~\cite{cheng2018model}, weights quantization and weights/feature maps sparsification~\cite{verelst2020dynamic}, and dynamic pruning~\cite{lin2017runtime}. While these strategies reduce the computation load at the edge, our methodology for canvas-based processing employs \emph{input pre-processing} to accelerate the concurrent processing of multiple multimedia sensor streams, and can be complemented by such DNN modification techniques. 

%% file: 08-conclusion.tex
\section{Conclusion}
\label{sec: conclusion}
We have introduced \name, a criticality-aware ``spatial multiplexing" based approach for DNN-based inferencing on edge devices that extracts high-priority regions of individual images and then spatially packs them into a composite canvas frame of smaller size, so as to ensure high processing throughput. \namef key innovation is the Mosaic-of-Scales (MoS) concept, a multi-scale tiling approach that ensures that objects of varying sizes are both represented at adequate dimensions, and with minimal redundancy on the canvas. Experimental studies with a representative Jetson TX2 edge device demonstrate how \names can provide a multiplicative increase in throughput—e.g., by packing critical regions from 6 distinct camera images into a single canvas frame, \names can achieve a cumulative throughput of $\sim$ 138 FPS while achieving pedestrian detection accuracy of 79\% on the Okutama-Action dataset. In contrast, processing each image frame individually provides a slight increase ($\le1\%$) in accuracy to 80\% but with sharply lower throughput ($\sim3FPS$), while simplistically packing 6 image frames uniformly into a canvas frame can achieve similar throughput ($\sim$144 FPS) but with significantly lower accuracy (71\%). Similar gains are observed for a separate LPR application, thereby demonstrating the generalizability of \name. In future work, we plan to complement \namef spatial packing strategy with additional \emph{temporal} packing, where high-priority regions from consecutive frames are extracted and packed on a single canvas.


%% file: main.bbl

\begin{thebibliography}{37}


\ifx \showCODEN    \undefined \def \showCODEN     #1{\unskip}     \fi
\ifx \showDOI      \undefined \def \showDOI       #1{#1}\fi
\ifx \showISBNx    \undefined \def \showISBNx     #1{\unskip}     \fi
\ifx \showISBNxiii \undefined \def \showISBNxiii  #1{\unskip}     \fi
\ifx \showISSN     \undefined \def \showISSN      #1{\unskip}     \fi
\ifx \showLCCN     \undefined \def \showLCCN      #1{\unskip}     \fi
\ifx \shownote     \undefined \def \shownote      #1{#1}          \fi
\ifx \showarticletitle \undefined \def \showarticletitle #1{#1}   \fi
\ifx \showURL      \undefined \def \showURL       {\relax}        \fi
\providecommand\bibfield[2]{#2}
\providecommand\bibinfo[2]{#2}
\providecommand\natexlab[1]{#1}
\providecommand\showeprint[2][]{arXiv:#2}

\bibitem[Abdelzaher et~al\mbox{.}(2020)]%
        {abdelzaher2020five}
\bibfield{author}{\bibinfo{person}{Tarek Abdelzaher}, \bibinfo{person}{Yifan
  Hao}, \bibinfo{person}{Kasthuri Jayarajah}, \bibinfo{person}{Archan Misra},
  \bibinfo{person}{Per Skarin}, \bibinfo{person}{Shuochao Yao},
  \bibinfo{person}{Dulanga Weerakoon}, {and} \bibinfo{person}{Karl-Erik
  {\AA}rz{\'e}n}.} \bibinfo{year}{2020}\natexlab{}.
\newblock \showarticletitle{Five challenges in cloud-enabled intelligence and
  control}.
\newblock \bibinfo{journal}{\emph{ACM Transactions on Internet Technology
  (TOIT)}} \bibinfo{volume}{20}, \bibinfo{number}{1} (\bibinfo{year}{2020}),
  \bibinfo{pages}{1--19}.
\newblock


\bibitem[Alon et~al\mbox{.}(2009)]%
        {alon2009online}
\bibfield{author}{\bibinfo{person}{Noga Alon}, \bibinfo{person}{Baruch
  Awerbuch}, \bibinfo{person}{Yossi Azar}, \bibinfo{person}{Niv Buchbinder},
  {and} \bibinfo{person}{Joseph Naor}.} \bibinfo{year}{2009}\natexlab{}.
\newblock \showarticletitle{The online set cover problem}.
\newblock \bibinfo{journal}{\emph{SIAM J. Comput.}} \bibinfo{volume}{39},
  \bibinfo{number}{2} (\bibinfo{year}{2009}), \bibinfo{pages}{361--370}.
\newblock


\bibitem[Barekatain et~al\mbox{.}(2017)]%
        {barekatain2017okutama}
\bibfield{author}{\bibinfo{person}{Mohammadamin Barekatain},
  \bibinfo{person}{Miquel Mart{\'\i}}, \bibinfo{person}{Hsueh-Fu Shih},
  \bibinfo{person}{Samuel Murray}, \bibinfo{person}{Kotaro Nakayama},
  \bibinfo{person}{Yutaka Matsuo}, {and} \bibinfo{person}{Helmut Prendinger}.}
  \bibinfo{year}{2017}\natexlab{}.
\newblock \showarticletitle{Okutama-action: An aerial view video dataset for
  concurrent human action detection}. In \bibinfo{booktitle}{\emph{Proceedings
  of the IEEE conference on computer vision and pattern recognition
  workshops}}. \bibinfo{pages}{28--35}.
\newblock


\bibitem[Bateni and Liu(2018)]%
        {bateni2018apnet}
\bibfield{author}{\bibinfo{person}{Soroush Bateni} {and} \bibinfo{person}{Cong
  Liu}.} \bibinfo{year}{2018}\natexlab{}.
\newblock \showarticletitle{Apnet: Approximation-aware real-time neural
  network}. In \bibinfo{booktitle}{\emph{2018 IEEE Real-Time Systems Symposium
  (RTSS)}}. IEEE, \bibinfo{pages}{67--79}.
\newblock


\bibitem[Bateni et~al\mbox{.}(2020)]%
        {bateni2020co}
\bibfield{author}{\bibinfo{person}{Soroush Bateni}, \bibinfo{person}{Zhendong
  Wang}, \bibinfo{person}{Yuankun Zhu}, \bibinfo{person}{Yang Hu}, {and}
  \bibinfo{person}{Cong Liu}.} \bibinfo{year}{2020}\natexlab{}.
\newblock \showarticletitle{Co-Optimizing Performance and Memory Footprint Via
  Integrated CPU/GPU Memory Management, an Implementation on Autonomous Driving
  Platform}. In \bibinfo{booktitle}{\emph{2020 IEEE Real-Time and Embedded
  Technology and Applications Symposium (RTAS)}}. IEEE,
  \bibinfo{pages}{310--323}.
\newblock


\bibitem[Bhattacharya and Lane(2016)]%
        {bhattacharya-sensys16}
\bibfield{author}{\bibinfo{person}{Sourav Bhattacharya} {and}
  \bibinfo{person}{Nicholas~D. Lane}.} \bibinfo{year}{2016}\natexlab{}.
\newblock \showarticletitle{Sparsification and Separation of Deep Learning
  Layers for Constrained Resource Inference on Wearables}. In
  \bibinfo{booktitle}{\emph{Proceedings of the 14th ACM Conference on Embedded
  Network Sensor Systems CD-ROM}} \emph{(\bibinfo{series}{SenSys '16})}.
  \bibinfo{publisher}{Association for Computing Machinery},
  \bibinfo{pages}{176–189}.
\newblock


\bibitem[Bolukbasi et~al\mbox{.}(2017)]%
        {bolukbasi2017adaptive}
\bibfield{author}{\bibinfo{person}{Tolga Bolukbasi}, \bibinfo{person}{Joseph
  Wang}, \bibinfo{person}{Ofer Dekel}, {and} \bibinfo{person}{Venkatesh
  Saligrama}.} \bibinfo{year}{2017}\natexlab{}.
\newblock \showarticletitle{Adaptive neural networks for efficient inference}.
  In \bibinfo{booktitle}{\emph{International Conference on Machine Learning}}.
  PMLR, \bibinfo{pages}{527--536}.
\newblock


\bibitem[Chen et~al\mbox{.}(2015)]%
        {chen2015glimpse}
\bibfield{author}{\bibinfo{person}{Tiffany Yu-Han Chen}, \bibinfo{person}{Lenin
  Ravindranath}, \bibinfo{person}{Shuo Deng}, \bibinfo{person}{Paramvir Bahl},
  {and} \bibinfo{person}{Hari Balakrishnan}.} \bibinfo{year}{2015}\natexlab{}.
\newblock \showarticletitle{Glimpse: Continuous, real-time object recognition
  on mobile devices}. In \bibinfo{booktitle}{\emph{Proceedings of the 13th ACM
  Conference on Embedded Networked Sensor Systems}}. \bibinfo{pages}{155--168}.
\newblock


\bibitem[Cheng et~al\mbox{.}(2018)]%
        {cheng2018model}
\bibfield{author}{\bibinfo{person}{Yu Cheng}, \bibinfo{person}{Duo Wang},
  \bibinfo{person}{Pan Zhou}, {and} \bibinfo{person}{Tao Zhang}.}
  \bibinfo{year}{2018}\natexlab{}.
\newblock \showarticletitle{Model compression and acceleration for deep neural
  networks: The principles, progress, and challenges}.
\newblock \bibinfo{journal}{\emph{IEEE Signal Processing Magazine}}
  \bibinfo{volume}{35}, \bibinfo{number}{1} (\bibinfo{year}{2018}),
  \bibinfo{pages}{126--136}.
\newblock


\bibitem[Chung and Park(2015)]%
        {chung2015notes}
\bibfield{author}{\bibinfo{person}{Yerim Chung} {and}
  \bibinfo{person}{Myoung-Ju Park}.} \bibinfo{year}{2015}\natexlab{}.
\newblock \showarticletitle{Notes on inverse bin-packing problems}.
\newblock \bibinfo{journal}{\emph{Inform. Process. Lett.}}
  \bibinfo{volume}{115}, \bibinfo{number}{1} (\bibinfo{year}{2015}),
  \bibinfo{pages}{60--68}.
\newblock


\bibitem[Gao et~al\mbox{.}(2018)]%
        {gao:cvpr2018}
\bibfield{author}{\bibinfo{person}{M. Gao}, \bibinfo{person}{R. Yu},
  \bibinfo{person}{A. Li}, \bibinfo{person}{V.~I. Morariu}, {and}
  \bibinfo{person}{L.~S. Davis}.} \bibinfo{year}{2018}\natexlab{}.
\newblock \showarticletitle{Dynamic Zoom-in Network for Fast Object Detection
  in Large Images}. In \bibinfo{booktitle}{\emph{2018 IEEE/CVF Conference on
  Computer Vision and Pattern Recognition (CVPR)}}. \bibinfo{publisher}{IEEE
  Computer Society}, \bibinfo{address}{Los Alamitos, CA, USA},
  \bibinfo{pages}{6926--6935}.
\newblock


\bibitem[Hsieh et~al\mbox{.}(2018)]%
        {hsieh2018focus}
\bibfield{author}{\bibinfo{person}{Kevin Hsieh}, \bibinfo{person}{Ganesh
  Ananthanarayanan}, \bibinfo{person}{Peter Bodik}, \bibinfo{person}{Shivaram
  Venkataraman}, \bibinfo{person}{Paramvir Bahl}, \bibinfo{person}{Matthai
  Philipose}, \bibinfo{person}{Phillip~B Gibbons}, {and} \bibinfo{person}{Onur
  Mutlu}.} \bibinfo{year}{2018}\natexlab{}.
\newblock \showarticletitle{Focus: Querying large video datasets with low
  latency and low cost}. In \bibinfo{booktitle}{\emph{13th USENIX Symposium on
  Operating Systems Design and Implementation (OSDI 18)}}.
  \bibinfo{pages}{269--286}.
\newblock


\bibitem[Hu et~al\mbox{.}(2021)]%
        {hu2021exploring}
\bibfield{author}{\bibinfo{person}{Yigong Hu}, \bibinfo{person}{Shengzhong
  Liu}, \bibinfo{person}{Tarek Abdelzaher}, \bibinfo{person}{Maggie Wigness},
  {and} \bibinfo{person}{Philip David}.} \bibinfo{year}{2021}\natexlab{}.
\newblock \showarticletitle{On exploring image resizing for optimizing
  criticality-based machine perception}. In \bibinfo{booktitle}{\emph{2021 IEEE
  27th International Conference on Embedded and Real-Time Computing Systems and
  Applications (RTCSA)}}. IEEE, \bibinfo{pages}{169--178}.
\newblock


\bibitem[Jiang et~al\mbox{.}(2018)]%
        {jiang2018chameleon}
\bibfield{author}{\bibinfo{person}{Junchen Jiang}, \bibinfo{person}{Ganesh
  Ananthanarayanan}, \bibinfo{person}{Peter Bodik}, \bibinfo{person}{Siddhartha
  Sen}, {and} \bibinfo{person}{Ion Stoica}.} \bibinfo{year}{2018}\natexlab{}.
\newblock \showarticletitle{Chameleon: scalable adaptation of video analytics}.
  In \bibinfo{booktitle}{\emph{Proceedings of the 2018 Conference of the ACM
  Special Interest Group on Data Communication}}. \bibinfo{pages}{253--266}.
\newblock


\bibitem[Jiang et~al\mbox{.}(2021)]%
        {jiang2021flexible}
\bibfield{author}{\bibinfo{person}{Shiqi Jiang}, \bibinfo{person}{Zhiqi Lin},
  \bibinfo{person}{Yuanchun Li}, \bibinfo{person}{Yuanchao Shu}, {and}
  \bibinfo{person}{Yunxin Liu}.} \bibinfo{year}{2021}\natexlab{}.
\newblock \showarticletitle{Flexible high-resolution object detection on edge
  devices with tunable latency}. In \bibinfo{booktitle}{\emph{Proceedings of
  the 27th Annual International Conference on Mobile Computing and
  Networking}}. \bibinfo{pages}{559--572}.
\newblock


\bibitem[Kang et~al\mbox{.}(2022)]%
        {kang2022dnn}
\bibfield{author}{\bibinfo{person}{Woosung Kang}, \bibinfo{person}{Siwoo
  Chung}, \bibinfo{person}{Jeremy~Yuhyun Kim}, \bibinfo{person}{Youngmoon Lee},
  \bibinfo{person}{Kilho Lee}, \bibinfo{person}{Jinkyu Lee},
  \bibinfo{person}{Kang~G Shin}, {and} \bibinfo{person}{Hoon~Sung Chwa}.}
  \bibinfo{year}{2022}\natexlab{}.
\newblock \showarticletitle{DNN-SAM: Split-and-Merge DNN Execution for
  Real-Time Object Detection}.
\newblock  (\bibinfo{date}{May} \bibinfo{year}{2022}).
\newblock


\bibitem[Kim et~al\mbox{.}(2020)]%
        {kim2020anytimenet}
\bibfield{author}{\bibinfo{person}{Jung-Eun Kim}, \bibinfo{person}{Richard
  Bradford}, {and} \bibinfo{person}{Zhong Shao}.}
  \bibinfo{year}{2020}\natexlab{}.
\newblock \showarticletitle{Anytimenet: Controlling time-quality tradeoffs in
  deep neural network architectures}. In \bibinfo{booktitle}{\emph{2020 Design,
  Automation \& Test in Europe Conference \& Exhibition (DATE)}}. IEEE,
  \bibinfo{pages}{945--950}.
\newblock


\bibitem[Laroca et~al\mbox{.}(2018)]%
        {laroca2018robust}
\bibfield{author}{\bibinfo{person}{Rayson Laroca}, \bibinfo{person}{Evair
  Severo}, \bibinfo{person}{Luiz~A Zanlorensi}, \bibinfo{person}{Luiz~S
  Oliveira}, \bibinfo{person}{Gabriel~Resende Gon{\c{c}}alves},
  \bibinfo{person}{William~Robson Schwartz}, {and} \bibinfo{person}{David
  Menotti}.} \bibinfo{year}{2018}\natexlab{}.
\newblock \showarticletitle{A robust real-time automatic license plate
  recognition based on the YOLO detector}. In \bibinfo{booktitle}{\emph{2018
  international joint conference on neural networks (ijcnn)}}. IEEE,
  \bibinfo{pages}{1--10}.
\newblock


\bibitem[Lin et~al\mbox{.}(2017)]%
        {lin2017runtime}
\bibfield{author}{\bibinfo{person}{Ji Lin}, \bibinfo{person}{Yongming Rao},
  \bibinfo{person}{Jiwen Lu}, {and} \bibinfo{person}{Jie Zhou}.}
  \bibinfo{year}{2017}\natexlab{}.
\newblock \showarticletitle{Runtime neural pruning}.
\newblock \bibinfo{journal}{\emph{Advances in neural information processing
  systems}}  \bibinfo{volume}{30} (\bibinfo{year}{2017}).
\newblock


\bibitem[Lin et~al\mbox{.}(2014)]%
        {lin2014microsoft}
\bibfield{author}{\bibinfo{person}{Tsung-Yi Lin}, \bibinfo{person}{Michael
  Maire}, \bibinfo{person}{Serge Belongie}, \bibinfo{person}{James Hays},
  \bibinfo{person}{Pietro Perona}, \bibinfo{person}{Deva Ramanan},
  \bibinfo{person}{Piotr Doll{\'a}r}, {and} \bibinfo{person}{C~Lawrence
  Zitnick}.} \bibinfo{year}{2014}\natexlab{}.
\newblock \showarticletitle{Microsoft coco: Common objects in context}. In
  \bibinfo{booktitle}{\emph{European conference on computer vision}}. Springer,
  \bibinfo{pages}{740--755}.
\newblock


\bibitem[Liu et~al\mbox{.}(2022)]%
        {liu2022self}
\bibfield{author}{\bibinfo{person}{Shengzhong Liu}, \bibinfo{person}{Xinzhe
  Fu}, \bibinfo{person}{Maggie Wigness}, \bibinfo{person}{Philip David},
  \bibinfo{person}{Shuochao Yao}, \bibinfo{person}{Liu Sha}, {and}
  \bibinfo{person}{Tarek Abdelzaher}.} \bibinfo{year}{2022}\natexlab{}.
\newblock \showarticletitle{Self-Cueing Real-Time Attention Scheduling in
  Criticality-Driven Visual Machine Perception}. In
  \bibinfo{booktitle}{\emph{In Proc. 28th IEEE Real-Time and Embedded
  Technology and Applications Symposium (RTAS)}}.
\newblock


\bibitem[Liu et~al\mbox{.}(2020)]%
        {liu2020removing}
\bibfield{author}{\bibinfo{person}{Shengzhong Liu}, \bibinfo{person}{Shuochao
  Yao}, \bibinfo{person}{Xinzhe Fu}, \bibinfo{person}{Rohan Tabish},
  \bibinfo{person}{Simon Yu}, \bibinfo{person}{Heechul Yun},
  \bibinfo{person}{Lui Sha}, {and} \bibinfo{person}{Tarek Abdelzaher}.}
  \bibinfo{year}{2020}\natexlab{}.
\newblock \showarticletitle{On Removing Algorithmic Priority Inversion from
  Mission-critical Machine Inference Pipelines}.
\newblock  (\bibinfo{date}{December} \bibinfo{year}{2020}).
\newblock


\bibitem[Liu et~al\mbox{.}(2016)]%
        {liu2016ssd}
\bibfield{author}{\bibinfo{person}{Wei Liu}, \bibinfo{person}{Dragomir
  Anguelov}, \bibinfo{person}{Dumitru Erhan}, \bibinfo{person}{Christian
  Szegedy}, \bibinfo{person}{Scott Reed}, \bibinfo{person}{Cheng-Yang Fu},
  {and} \bibinfo{person}{Alexander~C Berg}.} \bibinfo{year}{2016}\natexlab{}.
\newblock \showarticletitle{Ssd: Single shot multibox detector}. In
  \bibinfo{booktitle}{\emph{European conference on computer vision}}. Springer,
  \bibinfo{pages}{21--37}.
\newblock


\bibitem[McKerns et~al\mbox{.}(2021)]%
        {mysticlib}
\bibfield{author}{\bibinfo{person}{M McKerns}, \bibinfo{person}{P Hung}, {and}
  \bibinfo{person}{M Aivazis}.} \bibinfo{year}{2021}\natexlab{}.
\newblock \bibinfo{title}{mystic: highly-constrained non-convex optimization
  and UQ}.
\newblock
\newblock
\urldef\tempurl%
\url{https://github.com/uqfoundation/mystic}
\showURL{%
\tempurl}


\bibitem[NVIDIA(2022a)]%
        {tx2}
\bibfield{author}{\bibinfo{person}{NVIDIA}.} \bibinfo{year}{2022}\natexlab{a}.
\newblock \showarticletitle{Jetson TX2 Developer Kit}.
\newblock \bibinfo{journal}{\emph{NVIDIA}} (\bibinfo{year}{2022}).
\newblock
\urldef\tempurl%
\url{https://developer.nvidia.com/embedded/jetson-tx2}
\showURL{%
\tempurl}


\bibitem[NVIDIA(2022b)]%
        {xavier}
\bibfield{author}{\bibinfo{person}{NVIDIA}.} \bibinfo{year}{2022}\natexlab{b}.
\newblock \showarticletitle{Jetson Xavier NX Developer Kit}.
\newblock \bibinfo{journal}{\emph{NVIDIA}} (\bibinfo{year}{2022}).
\newblock
\urldef\tempurl%
\url{https://developer.nvidia.com/embedded/jetson-xavier-nx-devkit}
\showURL{%
\tempurl}


\bibitem[Olmedo et~al\mbox{.}(2020)]%
        {olmedo2020dissecting}
\bibfield{author}{\bibinfo{person}{Ignacio~Sa{\~n}udo Olmedo},
  \bibinfo{person}{Nicola Capodieci}, \bibinfo{person}{Jorge~Luis Martinez},
  \bibinfo{person}{Andrea Marongiu}, {and} \bibinfo{person}{Marko Bertogna}.}
  \bibinfo{year}{2020}\natexlab{}.
\newblock \showarticletitle{Dissecting the CUDA scheduling hierarchy: a
  performance and predictability perspective}. In
  \bibinfo{booktitle}{\emph{2020 IEEE Real-Time and Embedded Technology and
  Applications Symposium (RTAS)}}. IEEE, \bibinfo{pages}{213--225}.
\newblock


\bibitem[Otterness and Anderson(2020)]%
        {otterness2020amd}
\bibfield{author}{\bibinfo{person}{Nathan Otterness} {and}
  \bibinfo{person}{James~H Anderson}.} \bibinfo{year}{2020}\natexlab{}.
\newblock \showarticletitle{AMD GPUs as an alternative to NVIDIA for supporting
  real-time workloads}. In \bibinfo{booktitle}{\emph{32nd Euromicro Conference
  on Real-Time Systems (ECRTS 2020)}}. Schloss Dagstuhl-Leibniz-Zentrum f{\"u}r
  Informatik.
\newblock


\bibitem[Stone et~al\mbox{.}(2019)]%
        {stone2019towards}
\bibfield{author}{\bibinfo{person}{Theodore Stone}, \bibinfo{person}{Nathaniel
  Stone}, \bibinfo{person}{Puneet Jain}, \bibinfo{person}{Yurong Jiang},
  \bibinfo{person}{Kyu-Han Kim}, {and} \bibinfo{person}{Srihari Nelakuditi}.}
  \bibinfo{year}{2019}\natexlab{}.
\newblock \showarticletitle{Towards scalable video analytics at the edge}. In
  \bibinfo{booktitle}{\emph{2019 16th Annual IEEE International Conference on
  Sensing, Communication, and Networking (SECON)}}. IEEE,
  \bibinfo{pages}{1--9}.
\newblock


\bibitem[Ultralytics(2022)]%
        {yolov5}
\bibfield{author}{\bibinfo{person}{Ultralytics}.}
  \bibinfo{year}{2022}\natexlab{}.
\newblock \showarticletitle{Ultralytics/yolov5: Yolov5 in PyTorch, ONNX,
  CoreML, and TFLite}.
\newblock \bibinfo{journal}{\emph{GitHub}} (\bibinfo{year}{2022}).
\newblock
\urldef\tempurl%
\url{https://github.com/ultralytics/yolov5}
\showURL{%
\tempurl}


\bibitem[Verelst and Tuytelaars(2020)]%
        {verelst2020dynamic}
\bibfield{author}{\bibinfo{person}{Thomas Verelst} {and} \bibinfo{person}{Tinne
  Tuytelaars}.} \bibinfo{year}{2020}\natexlab{}.
\newblock \showarticletitle{Dynamic convolutions: Exploiting spatial sparsity
  for faster inference}. In \bibinfo{booktitle}{\emph{Proceedings of the
  ieee/cvf conference on computer vision and pattern recognition}}.
  \bibinfo{pages}{2320--2329}.
\newblock


\bibitem[Wang et~al\mbox{.}(2018)]%
        {xin-eccv18}
\bibfield{author}{\bibinfo{person}{Xin Wang}, \bibinfo{person}{Fisher Yu},
  \bibinfo{person}{Zi-Yi Dou}, \bibinfo{person}{Trevor Darrell}, {and}
  \bibinfo{person}{Joseph~E. Gonzalez}.} \bibinfo{year}{2018}\natexlab{}.
\newblock \showarticletitle{SkipNet: Learning Dynamic Routing in Convolutional
  Networks}. In \bibinfo{booktitle}{\emph{Computer Vision – ECCV 2018: 15th
  European Conference}}. \bibinfo{publisher}{Springer-Verlag},
  \bibinfo{pages}{420–436}.
\newblock


\bibitem[Wu et~al\mbox{.}(2022)]%
        {wu2022mrim}
\bibfield{author}{\bibinfo{person}{Ji-Yan Wu}, \bibinfo{person}{Vithurson
  Subasharan}, \bibinfo{person}{Tuan Tran}, {and} \bibinfo{person}{Archan
  Misra}.} \bibinfo{year}{2022}\natexlab{}.
\newblock \showarticletitle{MRIM: Enabling Mixed-Resolution Imaging for
  Low-Power Pervasive Vision Tasks}. In \bibinfo{booktitle}{\emph{2022 IEEE
  International Conference on Pervasive Computing and Communications
  (PerCom)}}. IEEE, \bibinfo{pages}{44--53}.
\newblock


\bibitem[Yang et~al\mbox{.}(2022)]%
        {yang:infocom2022}
\bibfield{author}{\bibinfo{person}{K. Yang}, \bibinfo{person}{J. Yi},
  \bibinfo{person}{K. Lee}, {and} \bibinfo{person}{Y. Lee}.}
  \bibinfo{year}{2022}\natexlab{}.
\newblock \showarticletitle{FlexPatch: Fast and Accurate Object Detection for
  On-device High-Resolution Live Video Analytics}. In
  \bibinfo{booktitle}{\emph{IEEE INFOCOM}}. \bibinfo{publisher}{IEEE}.
\newblock


\bibitem[Yang(2020)]%
        {yukai_yang_2020_4294717}
\bibfield{author}{\bibinfo{person}{Yukai Yang}.}
  \bibinfo{year}{2020}\natexlab{}.
\newblock \bibinfo{booktitle}{\emph{{FastMOT: High-Performance Multiple Object
  Tracking Based on Deep SORT and KLT}}}.
\newblock
\urldef\tempurl%
\url{https://doi.org/10.5281/zenodo.4294717}
\showDOI{\tempurl}


\bibitem[Yi et~al\mbox{.}(2020)]%
        {yi:mobicom2020}
\bibfield{author}{\bibinfo{person}{Juheon Yi}, \bibinfo{person}{Sunghyun Choi},
  {and} \bibinfo{person}{Youngki Lee}.} \bibinfo{year}{2020}\natexlab{}.
\newblock \showarticletitle{EagleEye: Wearable Camera-Based Person
  Identification in Crowded Urban Spaces}. In
  \bibinfo{booktitle}{\emph{Proceedings of the 26th Annual International
  Conference on Mobile Computing and Networking}}.
  \bibinfo{publisher}{Association for Computing Machinery},
  \bibinfo{address}{New York, NY, USA}, Article \bibinfo{articleno}{4},
  \bibinfo{numpages}{14}~pages.
\newblock


\bibitem[Zhang et~al\mbox{.}(2021)]%
        {zhang2021elf}
\bibfield{author}{\bibinfo{person}{Wuyang Zhang}, \bibinfo{person}{Zhezhi He},
  \bibinfo{person}{Luyang Liu}, \bibinfo{person}{Zhenhua Jia},
  \bibinfo{person}{Yunxin Liu}, \bibinfo{person}{Marco Gruteser},
  \bibinfo{person}{Dipankar Raychaudhuri}, {and} \bibinfo{person}{Yanyong
  Zhang}.} \bibinfo{year}{2021}\natexlab{}.
\newblock \showarticletitle{Elf: accelerate high-resolution mobile deep vision
  with content-aware parallel offloading}. In
  \bibinfo{booktitle}{\emph{Proceedings of the 27th Annual International
  Conference on Mobile Computing and Networking}}. \bibinfo{pages}{201--214}.
\newblock


\end{thebibliography}
